\documentclass[conference]{IEEEtran}

\usepackage{cite}
\usepackage{amsmath,amssymb,amsfonts}
\usepackage{algorithmic}
\usepackage{graphicx}
\usepackage[T1]{fontenc}
\usepackage{textcomp}
\usepackage{xcolor}

\usepackage{csvsimple}
\usepackage{booktabs}
\usepackage{multirow}
\usepackage{colortbl}
\usepackage{tikz}
\usepackage[font=footnotesize]{caption}
\usepackage{subcaption}
\usepackage{hyperref}
\usepackage{verbatim}
\usepackage{listings}
\usepackage{tablefootnote}
\usepackage{subcaption}

\def\BibTeX{{\rm B\kern-.05em{\sc i\kern-.025em b}\kern-.08em
    T\kern-.1667em\lower.7ex\hbox{E}\kern-.125emX}}

\usepackage{color}
\definecolor{lightgray}{rgb}{0.95, 0.95, 0.95}
\definecolor{darkgray}{rgb}{0.4, 0.4, 0.4}
\definecolor{editorGray}{rgb}{0.95, 0.95, 0.95}
\definecolor{editorOcher}{rgb}{1, 0.5, 0} %
\definecolor{editorGreen}{rgb}{0, 0.5, 0} %
\definecolor{orange}{rgb}{1,0.45,0.13}		
\definecolor{olive}{rgb}{0.17,0.59,0.20}
\definecolor{brown}{rgb}{0.69,0.31,0.31}
\definecolor{purple}{rgb}{0.38,0.18,0.81}
\definecolor{lightblue}{rgb}{0.1,0.57,0.7}
\definecolor{lightred}{rgb}{1,0.4,0.5}
\usepackage{upquote}
\usepackage{listings}
\lstdefinelanguage{CSS}{
  keywords={color,background-image:,margin,padding,font,weight,display,position,top,left,right,bottom,list,style,border,size,white,space,min,width, transition:, transform:, transition-property, transition-duration, transition-timing-function},	
  sensitive=true,
  morecomment=[l]{//},
  morecomment=[s]{/*}{*/},
  morestring=[b]',
  morestring=[b]",
  alsoletter={:},
  alsodigit={-}
}

\lstdefinelanguage{JavaScript}{
  morekeywords={typeof, new, true, false, catch, function, return, null, catch, switch, var, if, in, while, do, else, case, break, const, let, async, await},
  morecomment=[s]{/*}{*/},
  morecomment=[l]//,
  morestring=[b]",
  morestring=[b]',
  morestring=[b]`
}

\lstdefinelanguage{HTML5}{
  language=html,
  sensitive=true,	
  alsoletter={<>=-},	
  morecomment=[s]{<!-}{-->},
  tag=[s],
  otherkeywords={
  >,
	<!DOCTYPE,
  </html, <html, <head, <title, </title, <style, </style, <link, </head, <meta, />,
	</body, <body,
	</div, <div, </div>, 
	</p, <p, </p>,
	</script, <script,
  <canvas, /canvas>, <svg, <rect, <animateTransform, </rect>, </svg>, <video, <source, <iframe, </iframe>, </video>, <image, </image>, <header, </header, <article, </article
  },
  ndkeywords={
  =,
  charset=, src=, id=, width=, height=, style=, type=, rel=, href=,
  fill=, attributeName=, begin=, dur=, from=, to=, poster=, controls=, x=, y=, repeatCount=, xlink:href=,
  margin:, padding:, background-image:, border:, top:, left:, position:, width:, height:, margin-top:, margin-bottom:, font-size:, line-height:,
  transform:, -moz-transform:, -webkit-transform:,
  animation:, -webkit-animation:,
  transition:,  transition-duration:, transition-property:, transition-timing-function:,
  }
}

\lstdefinestyle{htmlcssjs} {%
  basicstyle={\footnotesize\ttfamily},   
  frame=b,
  xleftmargin={0.75cm},
  numbers=left,
  stepnumber=1,
  firstnumber=1,
  numberfirstline=true,	
  identifierstyle=\color{black},
  keywordstyle=\color{blue}\bfseries,
  ndkeywordstyle=\color{editorGreen}\bfseries,
  stringstyle=\color{editorOcher}\ttfamily,
  commentstyle=\color{brown}\ttfamily,
  language=HTML5,
  alsolanguage=JavaScript,
  alsodigit={.:;},	
  tabsize=2,
  showtabs=false,
  showspaces=false,
  showstringspaces=false,
  extendedchars=true,
  breaklines=true,
  literate=%
  {Ö}{{\"O}}1
  {Ä}{{\"A}}1
  {Ü}{{\"U}}1
  {ß}{{\ss}}1
  {ü}{{\"u}}1
  {ä}{{\"a}}1
  {ö}{{\"o}}1
}
\lstdefinestyle{py} {%
language=python,
literate=%
*{0}{{{\color{lightred}0}}}1
{1}{{{\color{lightred}1}}}1
{2}{{{\color{lightred}2}}}1
{3}{{{\color{lightred}3}}}1
{4}{{{\color{lightred}4}}}1
{5}{{{\color{lightred}5}}}1
{6}{{{\color{lightred}6}}}1
{7}{{{\color{lightred}7}}}1
{8}{{{\color{lightred}8}}}1
{9}{{{\color{lightred}9}}}1,
basicstyle=\footnotesize\ttfamily, %
numbers=left,               %
numbersep=5pt,              %
tabsize=4,                  %
extendedchars=true,         %
breaklines=true,            %
keywordstyle=\color{blue}\bfseries,
frame=b,
commentstyle=\color{brown}\itshape,
stringstyle=\color{editorOcher}\ttfamily, %
showspaces=false,           %
showtabs=false,             %
xleftmargin=17pt,
framexleftmargin=17pt,
framexrightmargin=5pt,
framexbottommargin=4pt,
showstringspaces=false,      %
}%

\lstdefinestyle{json}{
    basicstyle=\ttfamily\small,
    columns=fullflexible,
    showstringspaces=false,
    commentstyle=\color{gray}\upshape,
    morestring=[b]",
    morestring=[b]',
    stringstyle=\color{blue},
    keywordstyle=\color{purple}\bfseries,
    frame=single,
    breaklines=true
}

\begin{document}

\title{Distributed Architecture Reconstruction of Polyglot and Multi-Repository Microservice Projects}

\author{
    \IEEEauthorblockN{1\textsuperscript{st} Oscar Manglaras}
    \IEEEauthorblockA{
        \textit{Adelaide University}\\
        Adelaide, Australia \\
        oscar.manglaras@adelaide.edu.au
    }
    \and
    \IEEEauthorblockN{2\textsuperscript{nd} Alex Farkas}
    \IEEEauthorblockA{
        \textit{Adelaide University}\\
        Adelaide, Australia \\
        alex.m.farkas@gmail.com
    }
    \and
    \IEEEauthorblockN{3\textsuperscript{rd} Thomas Woolford}
    \IEEEauthorblockA{%
        \textit{Swordfish Computing}\\
        Adelaide, Australia \\
        thomas.woolford@swordfish.com.au
    }
    \and
    \IEEEauthorblockN{4\textsuperscript{th} Christoph Treude}
    \IEEEauthorblockA{%
        \textit{Singapore Management University}\\
        Singapore \\
        ctreude@smu.edu.sg
    }
    \and
    \IEEEauthorblockN{5\textsuperscript{th} Markus Wagner}
    \IEEEauthorblockA{%
        \textit{Monash University}\\
        Clayton, Australia \\
        markus.wagner@monash.edu
    }
}

\author{
    \IEEEauthorblockN{
        Oscar Manglaras\IEEEauthorrefmark{1},
        Alex Farkas\IEEEauthorrefmark{2},
        Thomas Woolford\IEEEauthorrefmark{3},
        Christoph Treude\IEEEauthorrefmark{4}
        and Markus Wagner\IEEEauthorrefmark{5}
    }
    \IEEEauthorblockA{\IEEEauthorrefmark{1}
        \textit{Adelaide University}, Adelaide, Australia,
        oscar.manglaras@adelaide.edu.au
    }
    \IEEEauthorblockA{
        \IEEEauthorrefmark{2}
        \textit{Adelaide University}, Adelaide, Australia,
        alex.m.farkas@gmail.com
    }
    \IEEEauthorblockA{
        \IEEEauthorrefmark{3}
        \textit{Swordfish Computing}, Adelaide, Australia,
        thomas.woolford@swordfish.com.au
    }
    \IEEEauthorblockA{\IEEEauthorrefmark{4}
        \textit{Singapore Management University}, Singapore,
        ctreude@smu.edu.sg
    },
    \IEEEauthorblockA{\IEEEauthorrefmark{5}
        \textit{Monash University}, Clayton, Australia,
        markus.wagner@monash.edu
    }
}

\maketitle

\begin{abstract}

Microservice architectures encourage the use of small, independently developed services; however, this can lead to increased architectural complexity.
Accurate documentation is crucial, but is challenging to maintain due to the rapid, independent evolution of services.
While static architecture reconstruction provides a way to maintain up-to-date documentation, existing approaches suffer from technology limitations, mono-repo constraints, or high implementation barriers.
This paper presents a novel framework for static architecture reconstruction that supports technology-specific analysis modules, called \emph{extractors}, and supports \emph{distributed architecture reconstruction} in multi-repo environments.
We describe the core design concepts and algorithms that govern how extractors are executed, how data is passed between them, and how their outputs are unified.
Furthermore, the framework is interoperable with existing static analysis tools and algorithms,
allowing them to be invoked from or embedded within extractors.

\end{abstract}

\begin{IEEEkeywords}
Microservices, Architecture Reconstruction, Distributed Systems
\end{IEEEkeywords}

\section{Introduction}

Microservices are designed to be small and focused, each handling a specific
task~\cite{Duellmann2017a}. While this simplifies individual services,
it increases the number of services and shifts much of the system's
complexity to the communication layer~\cite{Aderaldo2017}. Keeping documentation
accurate and up-to-date in such environments can be challenging~\cite{Kleehaus2019},
especially as services can be developed independently and modified frequently.

Automating documentation generation can help keep documentation up-to-date~\cite{Robillard2017},
supporting tasks such as onboarding, development, and debugging. To do this effectively,
it is essential to collect accurate architectural information. Static analysis of
source code and configuration files offers a way to reconstruct the system's
architecture without needing to deploy or run the services.
However, existing static analysis approaches focus on analysing specific technologies,
such as Java Spring, and lack the generality required to work across diverse systems.
The exceptions, which are language-agnostic,
require abstraction layers which raise the cost to support new languages
beyond what may be desirable or practical,
e.g., the ReSSA approach~\cite{Schiewe2022}.

To address these limitations, we propose a modular, static analysis-based framework
for reconstructing microservice architectures. Our approach introduces reusable
static analysis code modules called \emph{extractors}, which can be independently
developed and imported to support arbitrary technologies or integrate
existing static analysis tools. This paper presents the design concepts and algorithms
that define how these disparate static analysis modules are run,
how data is passed between them, and how to unify their outputs.

Furthermore, we introduce a distributed approach to architecture reconstruction
to support projects that use multiple source repositories (multi-repo source control),
a scenario ignored in existing research.
Instead of analysing the entire architecture at once, each service can generate an
architecture model as part of its build process. These models can then be aggregated
to form a complete architectural view, preserving the loose coupling principle fundamental to microservices.
Our reconstruction approach is designed to facilitate this process.

We focus on static analysis because mature dynamic analysis
solutions, such as the OpenTelemetry ecosystem, already exist. No such solutions
appear to exist for static analysis.
While both approaches have distinct advantages, static analysis does not require
the system to be running, making it better suited for integration into development
workflows~\cite{Cerny2022b} and for analysing individual microservice repositories.

Our open source implementation of this approach is available online
under the name ModARO\footnote{\url{https://gitlab.com/swordfish-computing-group/modaro}}
(MODular Architecture Reconstruction Orchestrator).

\section{Existing Work} \label{section:litreview}
Existing automated reconstruction approaches for microservice
architectures can broadly be divided into two categories;
static analysis, which uses artefacts such as source code and configuration files;
and dynamic analysis, which typically uses data from network traffic,
log outputs, and trace data~\cite{Cerny2022a}.
Most existing research has focused on dynamic analysis~\cite{Engel2018,Mayer2018,Kleehaus2018},
or a hybrid approach that extracts some metadata
from the code repository, but gathers inter-service
communication information dynamically~\cite{Granchelli2017,Ma2018}.

There is research, such as~\cite{Bushong2021,Rahman2019}, which present approaches
for reconstructing dependency graphs of microservices entirely
through static analysis; however, these approaches
are designed only to work with Java Spring, which makes them
useless for architectures using other languages or software frameworks.

To support multiple languages and technologies, Ntentos~\cite{Ntentos2021}
proposes an approach with programmable 'detectors'; class-based entities
for parsing and validating the static artifacts of an application
which developers can write and re-use between projects to
support different technologies.
However, the focus of this approach is architecture verification, not reconstruction;
users must explicitly define the entities to search for, and must manually pass
data to and between detectors.

In pursuit of language-agnosticism, Schiewe et al.~\cite{Schiewe2022a}
developed the ReSSA approach, which allows the definition of
analysers that extract code structures from a language-agnostic
abstract syntax tree (LAAST).
The approach requires generating language-specific
concrete syntax trees (CSTs), then converting these into an equivalent
LAAST structure.
Although this approach theoretically supports any language,
supporting new languages requires the use or creation of a parser
that can create a CST for the language, and code to translate
that CST into the equivalent LAAST.
In addition, the translation from CST to LAAST is not a trivial process;
requiring the identification of any and all
language-specific structures and statements (ternaries, match/switch, goto, etc.)
and the conversion of them into equivalent LAAST constructs,
or the modification of the LAAST if a language's features are impossible
to translate.~\cite{Bushong2022a}.
We believe that an approach with a lower barrier of entry for new languages
and technologies is viable and may be of greater practical use across a wide
range of architectures.

\section{The Problems and Proposed Solutions} \label{section:reconstruction-overview}

The greatest challenge in developing a general-purpose approach for statically analysing microservice architectures is the wide variety of technologies used;
not only across different projects, but also between individual services within a single project. Microservices can be written in different programming languages, depend on different libraries and frameworks, and follow distinct coding conventions. Even with a syntax parser for a given language, the representation of architectural concepts, such as endpoints or service calls, can vary significantly depending on the codebase’s dependencies and conventions, making it impossible for a single static analysis algorithm to support all microservice projects. This obstacle has likely contributed to the popularity of dynamic analysis approaches which rely on behavioural scrutiny, such as IP packet monitoring, or tracing frameworks like OpenTelemetry\footnote{\url{https://opentelemetry.io}}.

To overcome this limitation, we present a framework that prioritises easy prototyping and integration
of multiple analysis algorithms, rather than relying on a single universal algorithm.
Our approach divides reconstruction logic into modular components called \emph{extractors}
(Section~\ref{section:extractors}), each targeting a specific technology, such as Docker Compose
or Java Spring. These extractors can be reused for services that share the same technology stack.
Our reconstruction algorithm (Section~\ref{section:algorithm}) specifies how extractors are executed
and how their outputs are merged into a unified architecture model. We also provide an API
(Section~\ref{section:api}) to assist with common analysis tasks, reducing the effort required
to implement new extractors.

A second limitation of existing approaches is the implicit assumption that the full source code of the system is available during reconstruction. This assumption breaks down in ``multi-repo'' projects, where services are maintained in separate repositories by independent teams. While it is possible to pull all repositories and reconstruct the architecture centrally, doing so negates many of the benefits of a multi-repo setup;
even minor changes to one service would require re-analysing the entire system, and reconstruction couldn’t be integrated into a single service’s CI/CD pipeline without knowing which other services will be deployed alongside it.
This undermines the loose-coupling principle that is a common characteristic of how microservice architectures are developed and deployed~\cite{Lewis2014}, and fails to support their compositional nature, where a service may serve as a reusable component across multiple systems.

To address this, we introduce the concept of \emph{distributed architecture reconstruction}. Each repository runs the reconstruction process independently as part of its CI/CD pipeline, producing a \emph{model file} that captures its architectural information.
These \emph{model files} follow a data format (Section~\ref{section:metamodel}) that supports aggregation, enabling them to be merged into a single architecture-level \emph{model}.
This approach mirrors the componentised nature of microservices, allowing reconstructed architectural data to be updated independently and aggregated alongside the services themselves.
In this approach, extractors operate without knowledge of remote services; however, meaningful inter-service relationships, such as HTTP requests, still need to be expressed. To support this, extractors can define links that are resolved retroactively after model aggregation, preventing tight coupling between services during static analysis.

\section{Concepts and Terminology} \label{section:terminology}
The \emph{model} is a JSON-formatted hierarchical data structure produced by our approach.
When the \emph{model} is saved to a file, we call it a \emph{model file}.

A \emph{model entity} is any JSON
object\footnote{\url{https://datatracker.ietf.org/doc/html/rfc8259\#section-4}},
(a key-value map) or its in-memory equivalent (such as a JavaScript object
in our implementation) that conforms to the \emph{model schema}. A \emph{Model entity}
nested inside another is called a \emph{sub-entity}.
The \emph{top-level model entity} refers to the root object of the entire JSON structure.

The \emph{model schema} defines the expected structure of the \emph{model},
the types of \emph{model entities} that exist, and the fields each type can possess.
While we have a basic schema for testing purposes, 
presenting a general-purpose schema is beyond the scope of this paper.

An \emph{entity type} refers to a specific schema
for an entity within the \emph{model}.
A \emph{model entity} is of a given type if it conforms to that schema.
For example, in our testing, a \emph{microservice entity} was a
\emph{model entity} with the field \texttt{\$TYPE} set to the string \texttt{microservice},
and the optional string fields \texttt{name}, \texttt{version}, and
\texttt{\$path}.

\emph{Extractors} are modular pieces of code, implemented as functions,
that inspect and potentially modify \emph{model entities} during reconstruction
based on information from the source code.

\section{Extractors} \label{section:extractors}

\subsection{Design Principles}

\begin{figure}
    \centering
\begin{lstlisting}[style=htmlcssjs,breaklines=true]
async function extractor(entity) {
  const glob = entity.$path + "/**/*.java";
  const javaFiles = await getPaths(glob);
  if (javaFiles.length > 0)
    entity.java = true;
  return entity;
}

register_extractor(extractor, {
    properties: {
        $TYPE: { const: 'microservice' },
        $path: { type: 'string' },
    },
    required: ['$TYPE', '$path'],
})
\end{lstlisting}
    \caption{
        Am example of an extractor that tests if the microservice source
        code contains any Java files. The \texttt{\$path} field in the microservice
        entity would have been set by another extractor.
        \texttt{getPaths} is an example of an API function (see Section~\ref{section:api}).
    }
    \label{fig:extractor-fn}
\end{figure}

Each extractor is a single function that accepts a \emph{model entity}
as input, modifies it, and returns the modified entity as output.

\figurename~\ref{fig:extractor-fn} shows an example of a very simple extractor
that sets the \texttt{java} field on the \emph{microservice entity} to \texttt{true}
if the microservice source code contains any Java files.
The extractor function must then be registered with the reconstruction algorithm.
In our implementation, this was done by passing the function to a 
\texttt{register\_extractor} API method.

The extractors are expected to have no global or persistent state; instead, the
behaviour of the extractor is defined entirely by the contents of the input \emph{model
entity}.
In this design, the \emph{model} itself acts as the sole interface through which data can be
passed between extractors.
In practice, the behaviour of the extractor will also
depend on the contents of the code repository being analysed; however, we assume that
the repository is not being modified during the reconstruction process.
The stateless design would theoretically
allow extractors to be executed in parallel or memoised, improving performance.

To further simplify the design of the extractors, we are providing an extra
guarantee that each extractor will only be run once for each \emph{model entity}.
This makes some operations safer, such as appending items to arrays.

Our intention is for extractors to have no restrictions on their scripting capabilities.
They may load third-party libraries and invoke external processes,
allowing the reconstruction framework to integrate existing tools
for syntax parsing and static analysis.  For example, we integrated the aforementioned ReSSA
approach~\cite{Schiewe2022} by using an extractor as a wrapper to invoke the
ReSSA tool on the codebase and insert its results into \emph{model entities}.
We acknowledge that this level of flexibility introduces potential security
risks, particularly when running extractors from third-party sources.
While these concerns fall outside the scope of this paper,
they may be worth exploring in future work.

\subsection{Extractor Schemas}

Extractors may operate on different types of entities within the \emph{model},
such as the \emph{top-level model entity}, individual
microservices, or only those microservices
that define specific optional fields like \texttt{\$path}.
To support this versatility, extractors register a schema
that specifies the structure of the \emph{model entities} they accept
as input. We adopt the JSON Schema
specification\footnote{\url{https://json-schema.org/}} for this
purpose, as it is a mature standard for
describing JSON data, which aligns with the format
of our architecture \emph{model}.
\figurename~\ref{fig:extractor-fn} shows such a schema being registered
alongside the extractor function.

\subsection{Internal Model Fields} \label{section:extractor-model-fields}
Some extractors need to share information with others,
e.g., identifying which directory in a mono-repo corresponds to each microservice.
Since extractors have no shared global state, this information must be passed through the \emph{model} itself.
To support this, we reserve certain fields in the \emph{model} specifically
for passing data between extractors. Because we cannot predict all
the types of data that different extractors might need, we allow arbitrary
data to be stored in any key that matches the regular expression
\texttt{\textasciicircum\textbackslash\$[a-z0-9\_]+\$}~\footnote{\url{https://regex101.com/r/UoFYHU/1}}.

\section{The Reconstruction Algorithm} \label{section:algorithm}
The reconstruction algorithm defines the execution flow of extractors and the process
for integrating their outputs into a unified \emph{model}.
\subsection{Triggering the Algorithm}

\begin{figure}
    \centering
    \includegraphics[width=\linewidth]{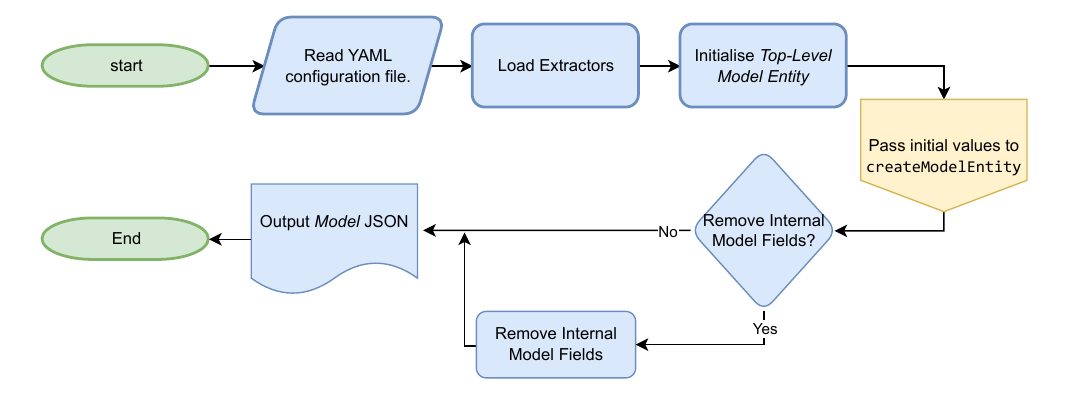}
    \caption{Initialisation process of the reconstruction approach.}
    \label{fig:initialisation}
\end{figure}

\begin{figure}
    \centering
    \includegraphics[width=\linewidth]{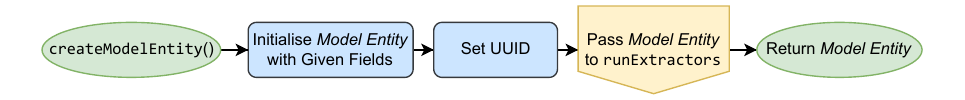}
    \caption{
        The \texttt{createModelEntity} function.
    }
    \label{fig:create-model-entity}
\end{figure}

The algorithm is triggered recursively; first, we create the \emph{top-level
model entity} with initial values
(e.g. the file path of the repository being analysed) defined in a
configuration file. The \texttt{createModelEntity} (\figurename~\ref{fig:create-model-entity}) function
is automatically called with these values, which calls \texttt{runExtractors} and begins
executing extractors designed for top-level entities
(\figurename~\ref{fig:initialisation}). These extractors identify and create new
\emph{microservice entities} using \texttt{createModelEntity},
which in turn triggers \texttt{runExtractors} to run microservice-specific extractors.
This recursive process continues
until no new entities are created or all applicable extractors
have been executed.

\subsection{The Algorithm}

\begin{figure}
    \centering
    \includegraphics[width=\linewidth]{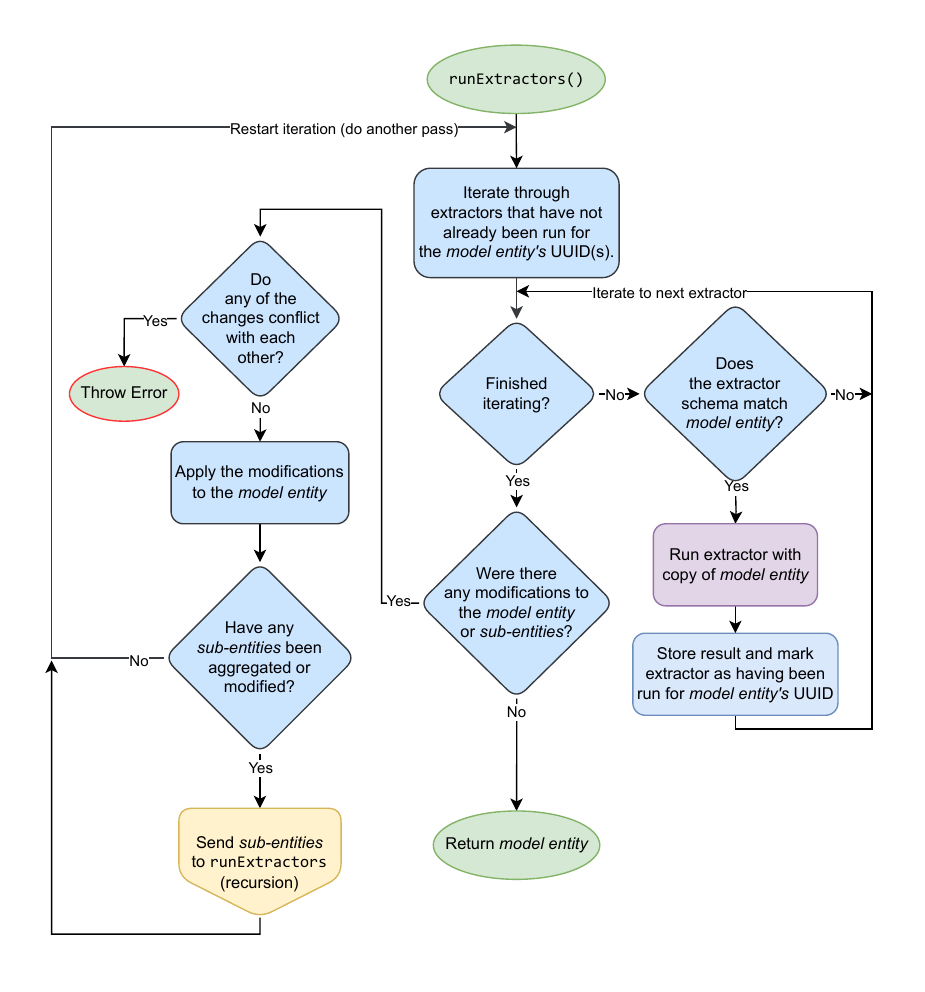}
    \caption{
        The algorithm for the \texttt{runExtractors} function.
        In the flowchart \emph{entity} refers to the \emph{model entity}
        passed to \texttt{runExtractors}.
    }
    \label{fig:run-extractors-simplified}
\end{figure}

The algorithm for \texttt{runExtractors} is shown in \figurename~\ref{fig:run-extractors-simplified}.
The algorithm iterates over the
list of registered extractors and tests if the \emph{model entity} conforms
to the extractor's schema. If so, a copy of the entity is sent to the
extractor, which returns the modified entity.
Once all extractors have finished running, we detect the changes made to the entity by
each extractor, apply them, and then do another pass over the list of
extractors to see if the now-modified entity conforms to any new
extractor schemas.
The entity is returned by \texttt{runExtractors} once a pass has been completed with no modifications.
Note that while \figurename~\ref{fig:run-extractors-simplified} shows the extractors
within a pass being run sequentially,
the algorithm is designed to allow them to run in parallel.

If the changes made to the \emph{model entity} in
the same pass conflict, then we throw an error.
A conflict occurs when two changes to an entity field are incompatible
according to the model aggregation
algorithm (see Section~\ref{section:aggregation}).
As per the aggregation algorithm, if two extractors
create non-scalar values - arrays or \emph{sub-entities} (key-value maps) - then the result
will be the union of those values, allowing multiple extractors to,
e.g., create and populate the same array with data collected from different sources.

Before starting a new pass, we first check if any of the \emph{sub-entities} within the
entity have been modified. This could occur if an extractor is making
changes to \emph{sub-entities}, or if the aggregation step created a union between two
entities.
In this scenario, we pass the modified \emph{sub-entities} to \texttt{runExtractors} again (recursion).
Only once this is complete do we continue with the next pass.

It is important to keep track of which extractors have been run on each
\emph{model entity}. We do not want the same extractors to be run on the same \emph{model entity}
multiple times as %
it could complicate the extractors
and easily lead to infinite loops (e.g., an extractor that pushes to an array).
A simple way to do this is
to give each \emph{model entity} a UUID when created (\figurename~\ref{fig:create-model-entity})
and record on which UUIDs an extractor has operated.
The UUID string is stored in an array within a reserved field;
if entities are aggregated, the UUID array then becomes a union of all UUIDs in the component entities,
precluding extractors that have run on any of them.

In theory, there may be cases where an extractor needs to run
multiple times on the same \emph{model entity}. For example,
an extractor modifying or building upon the results
of multiple other extractors. However, it is unclear whether such situations
would arise in practice. 
We could easily modify the algorithm to allow extractors to opt in to
repeatable execution if necessary; future work could investigate
whether this is required for practical applications.

\section{API Framework} \label{section:api}

To simplify the development of extractors, we provide an API for performing common operations.
Each extractor has direct access to this API.
During our testing, we identified three key operations that,
we posit, the majority of extractors need to perform.
Our API functions are designed to address
these operations. They are:
\begin{enumerate}
    \item [O1] Scanning the directory tree for specific files and folders.
    \item [O2] Extracting data from the contents of the files.
    \item [O3] Injecting data into the \emph{model}.
\end{enumerate}

APIs for injecting data into the \emph{model}, and the complexity or necessity of those APIs
will depend on the \emph{model schema},
e.g. a \texttt{createMicroservice} function which initialises and validates
the fields of a \emph{microservice entity}.
As the definition of a \emph{model schema} is out of scope for this paper, we
will not talk about model injection further.

Note that the API methods described here are not intended to
be final. The expectation is that new APIs would be created,
and new third-party libraries made available, as needed.

\subsection{Scanning the directory tree for valid files and folders}

There are two ways for an extractor to identify target files. The first is to
detect files that match specific filenames,
the second is to choose files that contain some specific text.
To match specific filenames, we chose to use the JavaScript \texttt{glob}
package\footnote{\url{https://www.npmjs.com/package/glob}}, which is an implementation
of the filename pattern expansion used by bash
shells\footnote{\url{https://www.gnu.org/software/bash/manual/html_node/Pattern-Matching.html}}.
We also provide other functions that detect files
containing regular expression matches.

\subsection{Extracting data from the contents of the file}
We expect that extracting data from source files will be the most complex task 
for extractors. While common formats like JSON, YAML, and XML can be parsed using
existing libraries, source code in languages like Java or Python poses greater
challenges. Tools like ANTLR\footnote{\url{https://www.antlr.org/}} exist, but
require full-language grammars that can be fragile to language
changes\footnote{\url{https://github.com/antlr/grammars-v4/issues}}.

To lower the implementation barrier for extractors, we instead
provide APIs that perform regular expression searches on the codebase.
We use the
Super Expressive\footnote{\url{https://github.com/francisrstokes/super-expressive}}
library, which simplifies composing and reusing complex expressions,
to provide pre-built expressions for common
matches like string literals and URLs.

This decision prioritises prototyping speed over parsing precision,
based on the hypothesis that much of the architectural information
developers care about can be extracted without comprehensive language parsing. 
However, our intention is to eventually provide both options,
offering APIs for full grammar parsing when needed,
while still allowing simpler extraction methods like RegEx when
they are sufficient.

\section{Distributed Architecture Reconstruction} \label{section:metamodel}

Earlier sections discuss how using JSON for the \emph{model} enables storage and transfer of arbitrary data
by and between extractors. This section focuses on the additional algorithms
and design decisions that allow the \emph{model} to facilitate distributed
architecture reconstruction; specifically,
how multiple \emph{models} (presumably in the form of \emph{model files})
can be aggregated into a unified \emph{model},
and how extractors can define links between \emph{model entities} that are resolved only after aggregation.

\subsection{Aggregation} \label{section:aggregation}

Aggregation is the act of combining multiple \emph{models} into one
while maintaining the same \emph{model schemas}.
For example, if two \emph{models} each contain a microservice
array with one entry, aggregation should result in the microservices
array containing both entries.
The aggregation algorithm
is also used by the reconstruction
algorithm (Section~\ref{section:algorithm}) to aggregate
the results of multiple extractors.

Since each \emph{model} is a JSON object, this task can be
framed as a recursive union of JSON objects. Our merging algorithm is
inspired
by~\cite{Kleppmann2017}, in which primitive types (strings, numbers,
booleans, and null) are treated as scalars, and objects are treated as maps
that can be merged recursively with a union operation.

Our approach differs in how it handles conflicts and arrays. When two
extractors make conflicting changes, such as assigning incompatible values
to the same key in the same object, we do not attempt to resolve the conflict
automatically. Instead, the algorithm raises an error, making the conflict
explicit and observable.
This ensures that the fields present in the input \emph{models} are
preserved in the aggregated output; fields are only ever added
as objects are aggregated,
never removed or modified.

Arrays are merged by forming a union of their contents. Values are considered
equivalent if they are compatible under the aggregation rules; i.e. identical
primitives or objects that can be merged without conflicts. Each item in the
first array is compared against items in the second: if an equivalent item is
found, the two are merged; if not, the item is added as a separate entry. As
a result, arrays of primitives contain one copy of each unique value, while
arrays of objects aggregate compatible entries. For example, \emph{microservice
entity} objects with matching names and version strings are merged, while distinct
ones remain separate.
This allows primitives, such as names, to serve as
implicit keys that define which entities should be merged.

\subsection{Retroactive Linking}

In a distributed reconstruction scenario, extractors operate on
microservices independently
and lack access to other parts of the system. As a result,
they may need to specify relationships, such as an HTTP request to another
microservice, without complete information about the target entity.
To support this scenario, our \emph{model} includes a mechanism for defining links to entities
that may not exist in the local reconstruction context but can be resolved
later, once \emph{model} have been aggregated.

Referencing objects in remote files is not a new concept;
OpenAPI allows such references using
the URI of the remote file\footnote{\url{https://spec.openapis.org/oas/v3.1.1.html\#reference-object}}.
Another approach is to use
string IDs that are resolved separately after the \emph{models} are
combined, similar to node anchors in
YAML\footnote{\url{https://yaml.org/spec/1.2.2/\#692-node-anchors}}.
However, approaches based on URIs or string IDs require that extractors
know specific details about external entities during reconstruction
and that those details remain stable over time. These assumptions
conflict with the principle of loose-coupling, which
distributed architecture reconstruction is intended to promote.

To avoid tight coupling, our approach expresses links in terms of observable
characteristics rather than fixed identifiers such as IDs or URLs. For
example, if an extractor detects a request to
\texttt{https://test-service:123/api/456}, the link can be defined as
targeting a microservice with domain \texttt{test-service}, port
\texttt{123}, and an endpoint at \texttt{/api/456}---without requiring
knowledge of the service’s repository, name, ID, or version.
Our implementation defines a \emph{model entity} with a \texttt{\$TYPE}
field set to \texttt{\$LINK}, and uses JSON Schema to
specify the link target and a
JSON Pointer\footnote{\url{https://doi.org/10.17487/RFC6901}}
to specify an array containing the target.

\section{Potential Applications} \label{section:applications}
Although our primary goal is to reconstruct microservice architectures
for the purpose of generating documentation, the flexibility of our
approach makes it applicable beyond the microservice domain.
Since extractors are modular and unconstrained in the data they can
produce, this framework can be used to collect and integrate nearly
any kind of static information from a codebase.

For example, Swordfish Computing (the company which co-funded this research)
has used our framework to generate
dependency graphs for their microservice projects. When aggregating
\emph{models} across services, the system automatically merges shared
dependencies (e.g., identical library versions) into a single entry,
making the resulting graph more compact and easier to interpret.

Other potential uses could be to run a suite of
heuristic and code smell tools within extractors and
collate all their outputs, to detect common configuration issues
and problematic idioms, and to run tests
as part of a build pipeline and collate the reports.

\section{Conclusion}

In this paper, we present a novel framework for reconstructing microservice architectures through static analysis,
with the ultimate goal of automatically generating architectural documentation.
Our framework splits reconstruction logic into independent modules called ``extractors,'' each tailored to
a specific technology.
The framework manages the invocation of extractors and the data flow between
them, provides an API for common operations,
and merges their outputs into a unified \emph{model}.
Existing static analysis algorithms or tools can be integrated
by implementing or calling them from an extractor.
While our focus is on microservice architectures,
we believe the approach is also applicable to other domains.

In addition, our design and model algorithms support the concept of distributed architecture reconstruction,
in which microservice repositories can be analysed independently and their results
aggregated later to form a complete architectural view.
This enables extractors operating on separate repositories to function with minimal knowledge of other services,
reducing coupling across microservices.

Future work includes defining a standard schema for the \emph{model};
exploring appropriate code-parsing tools or libraries that would be worth
providing as a standard API to extractors;
evaluating the framework on real-world projects;
and collecting feedback from practitioners.

\newpage
\bibliographystyle{IEEEtran}

\end{document}